\documentclass[12pt]{article}
\usepackage{graphicx} % Required for inserting images
\usepackage{comment}
\usepackage{caption}
\usepackage[utf8]{inputenc}
\usepackage{nameref}
\usepackage{array}
\newcolumntype{P}[1]{>{\centering\arraybackslash}p{#1}}
\usepackage{placeins}
\usepackage{todonotes}
\usepackage{subcaption}
\usepackage[percent]{overpic}
\usepackage{amsmath}
\usepackage[a4paper, margin=2.5cm]{geometry}
\usepackage[
  backend=biber,
  style=numeric-comp,
  sorting=none,
  giveninits=true,
  maxbibnames=3,
  minbibnames=3,
  doi=false,
  isbn=false,
  url=false
]{biblatex}
\usepackage{soul}
\usepackage{xcolor}
\sethlcolor{yellow!70}
\title{Detective Quantum Efficiency of the Timepix4 Hybrid Pixel Detector and its Application to Parallel-Beam Diffraction}
\author{Zhiyuan Ding $^{1,2}$, Nina Dimova $^3$,  Jonathan S. Barnard $^2$,\\ Giulio Crevatin $^4$, Liam O'Ryan $^4$, Richard Plackett $^3$,\\ Daniela Bortoletto $^3$, Angus I. Kirkland $^{1,2}$ and Marcus Gallagher-Jones $^2$\\
[10pt]
$^1$ University of Oxford, Department of Materials, Parks Road,\\ Oxford, OX1 3PH, UK\\
$^2$ The Rosalind Franklin Institute, Harwell Campus, \\Didcot, OX11 0QX, UK\\
$^3$ University of Oxford, Department of Physics, Parks Road,\\ Oxford OX1 3PU, UK\\
$^4$ Quantum Detectors Ltd., Harwell Campus,\\ Didcot, OX11 0QX, UK\\
}
\date{}

\begin{document}

\maketitle

\begin{abstract}

The detective quantum efficiency (DQE) and normalised noise power spectrum (NNPS) of the Timepix4 hybrid pixel detector in event-driven mode in TEM have been measured at 100~kV and 200~kV. In a raw data readout mode, the zero-frequency DQE exceeds 0.9 at both 100~kV and 200~kV. At the Nyquist frequency, the DQE remains above 0.2 at 100~kV but drops close to zero at 200~kV. Initial parallel-beam diffraction data from a polycrystalline gold nanoparticle sample is reported which shows that at 200~kV Timepix4 can detect weak diffracted information beyond a 75~mrad half-angle.

\end{abstract}

\section{Introduction}

The Timepix4 is the newest member of the Medipix family of readout ASICs~\cite{ballabriga2020introduction}. As an event-based hybrid pixel detector with 195~ps event binning~\cite{heijhoff2022timing,llopart2022timepix4}, it has been widely used for the detection of high energy particles~\cite{akiba2025timepix4}, X-rays~\cite{delogu2024validation,chushkin2025achieving,correa2024tempus} and photons from the visible to gamma rays~\cite{bolzonella2025development,bolzonella2024timing, cerbone2023monte}. In applications for transmission electron microscopy (TEM), the Timepix4 detector offers clear advantages for high timing resolution data recording. This is particularly important for data collection in various imaging geometries, including 4DSTEM~\cite{jannis2022event,li2025atomically,annys2025removing}, tomography~\cite{migunov2015rapid,koneti2019fast} and microED/3DED~\cite{gallagher2020atomic,vlahakis2025fast}, at low fluences which minimise radiation damage to the sample.

The Timepix4 detector offers an event-driven readout rate of up to $3.58$~Mhits/mm$^2$/s, corresponding to $1.12 \times 10^4$~events/pixel/s under uniform illumination. This represents a significant increase compared to earlier Medipix-family detectors, which are typically limited by frame-based readout architectures and / or lower bandwidths. For comparison, the Timepix3 detector is limited to  a readout rate of $3.9 \times 10^2$~events/pixel/s~\mbox{\cite{poikela2014timepix3}}.

A key architectural feature of Timepix4 is that the full $512 \times 448$ pixel matrix is implemented on a single ASIC. This eliminates the presence of inter-chip boundaries and associated edge pixels that are common in tiled detector systems based on smaller ASICs (e.g. $256 \times 256$ Medipix2 modules). As a result, the detector provides a larger continuous sensitive area without gaps, which is advantageous for quantitative measurements such as the MTF and DQE, as well as for diffraction and imaging applications requiring uniform spatial response.

For practical applications in TEM, the detector response as a function of the electron beam energy and fluence requires characterisation. Quantitative measurements including the modulation transfer function (MTF) and detective quantum efficiency (DQE) are representative metrics that characterise the detector’s response to electron hits~\cite{meyer1998effects,meyer2000experimental,meyer2000characterisation}. The MTF reflects the resolution of the detector as a function of spatial frequency~\cite{meyer2000experimental} and the DQE reflects the transfer of information above noise across different spatial frequencies~\cite{ruskin2013quantitative,mcmullan2009detective}.

Using Timepix4 for TEM, Dimova \textit{et al.}~\cite{dimova2025measurement} have previously measured the MTF at 100~kV and 200~kV acceleration voltages and improved the MTF using a clustering algorithm to reduce effects from charge sharing. Ghosh \textit{et al.}~\cite{ghosh2025timepix4}, in a technical report, have calculated the MTF and DQE at 300~kV using Monte-Carlo simulations of the electron trajectories in the sensor and discussed the influence of sensor thinning and different event threshold settings on the DQE and MTF.

In this work, the normalised noise power spectrum (NNPS) and DQE of a Timepix4 detector in TEM at 100~kV and 200~kV have been measured. The NNPS is obtained from flat-field illumination datasets and subsequently the DQE is calculated from the measured NNPS and MTF using a slanted knife-edge as a deterministic input signal~\cite{dimova2025measurement}. Initial experimental data using the Timepix4 detector to record data in a parallel-beam diffraction geometry from a sample of polycrystalline gold nanoparticles is also presented as an example application.

\FloatBarrier
\section{DQE Measurement}
\label{sec:DQE_measurement}

\subsection{Detector System}
\label{sec:detector_system}
The Timepix4 ASIC was bump-bonded to a $300~\mu$m planar silicon sensor. The detector assembly was installed on a JEOL CryoARM Z300FSC TEM with the Merlin T4 readout system, which provides $81.92$~Gbps FireFly optical readout. In event-driven mode, the maximum detection speed is approximately $1.79~$Mhits/mm$^2$/s, limited by the transfer bandwidth. The physical pixel size (pixel pitch) is $55~\mu$m and the number of pixels is $512 \times 448$. The theoretical maximum number of events that can be detected (calculated from the maximum detection speed) is approximately $5.58 \times 10 ^3$~events/pixel/s.

The per-pixel threshold for signal detection was set at $1000~$electron-hole pairs (approximately equivalent to the charge generated by 3.6~keV photon) in the sensor and a reverse-bias voltage of $100~$V was applied.

This per-pixel threshold was chosen based on two main considerations. First, a lower threshold increases the overall sensitivity of the detector. When combined with the high temporal resolution of the detector, this is particularly advantageous for fast, low-fluence imaging conditions, where maximizing fluence efficiency is critical.

Second, a lower threshold enables the detection of a larger fraction of charge-sharing events, resulting in more information-rich clusters. This is expected to facilitate more accurate cluster reconstruction, which provides possibilities for further detailed analysis in the future.

However, this work focuses on the intrinsic performance of the detector. Therefore, the MTF, NPS, and DQE calculation are based on raw data with no clustering applied. This provides a baseline against which subsequent reconstruction and calibration improvements can be evaluated.

\subsection{Methods}
\label{sec:method_DQE_and_NPS_cal}

\subsubsection{MTF Measurement}
\label{sec:methods_mtf_meas}

\begin{figure}[htbp]
    \centering
    \begin{subfigure}[t]{0.17\textwidth}
        \centering
        \includegraphics[width=\textwidth]{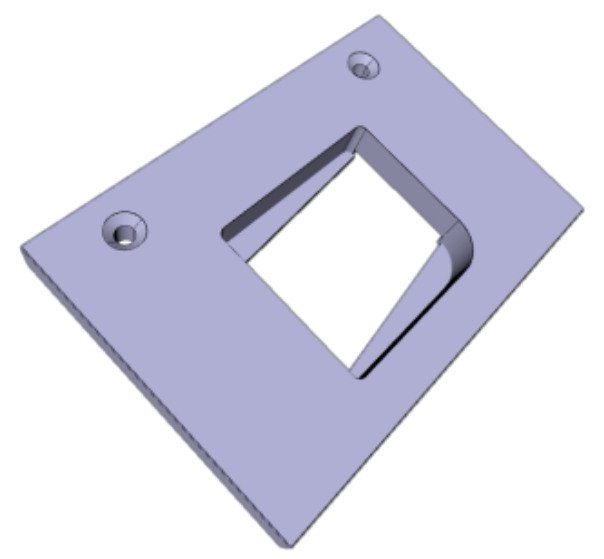}
        \caption{}
    \end{subfigure}
    \hspace{0.01\textwidth}
    \begin{subfigure}[t]{0.31\textwidth}
        \centering
        \includegraphics[width=\textwidth]{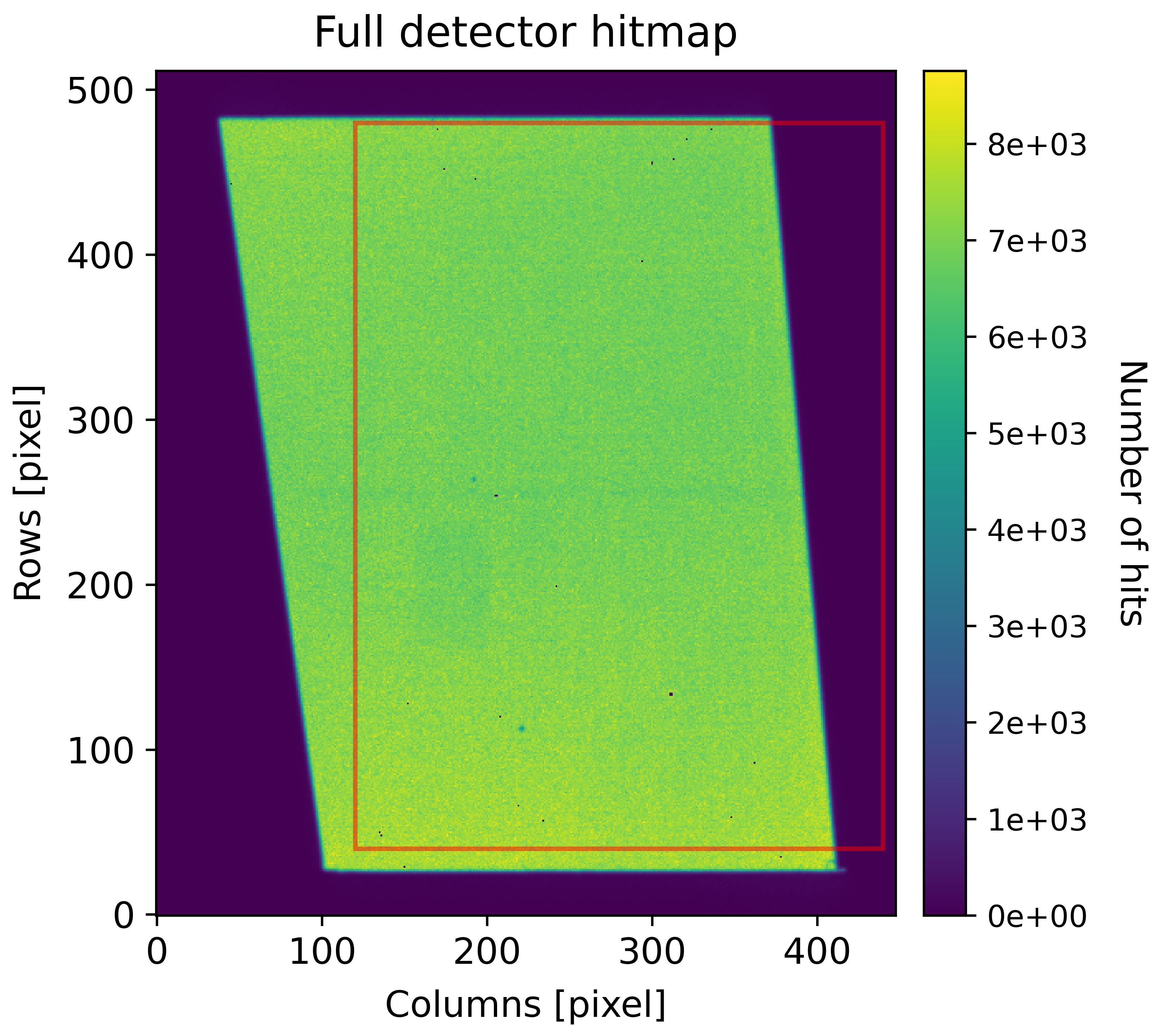}
        \caption{}
    \end{subfigure}
    \hspace{0.02\textwidth}
    \begin{subfigure}[t]{0.45\textwidth}
        \centering
        \includegraphics[width=\textwidth]{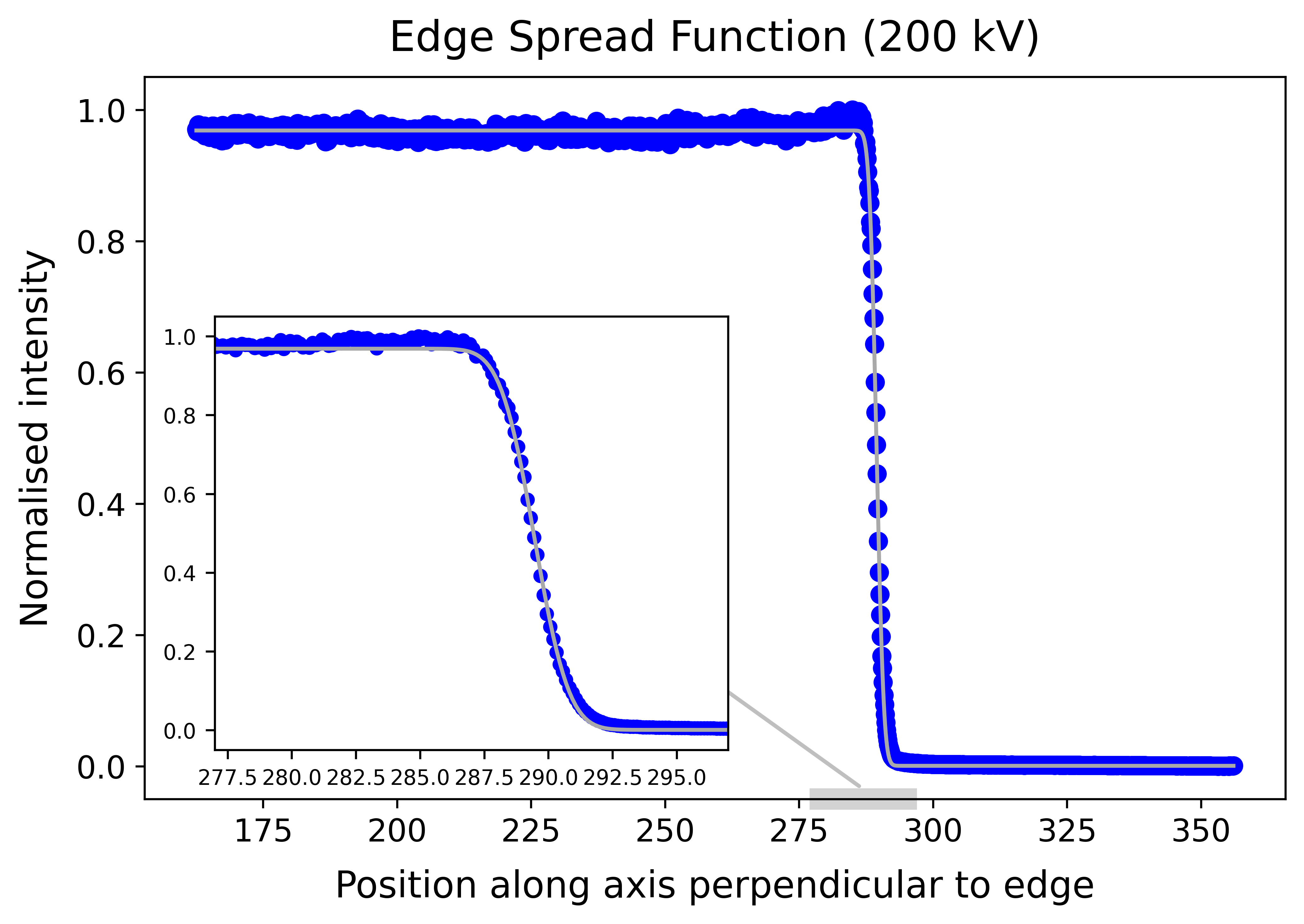}
        \caption{}
    \end{subfigure}
    \caption{(a) The custom aluminium knife-edge fabricated for electron data collection. It consists of a `plate' and two `steps'. The steps form angles $5^{\circ}$ and $8^{\circ}$ with respect to the long side of the plate. The plate is manufactured so that it is aligned parallel to the pixel matrix when attached to the detector casing. (b) The sampling region selected for MTF calculation, marked by a red outline. A large part of the bright area is used to facilitate averaging of the ESF. (c) The ESF measured at 200 kV and the fit function (grey) used to extract the MTF. Small deviations from an ideal monotonic step are observed on both sides of the transition. These features arise from the projection and binning process, which mixes contributions from pixels spanning a finite range of distances from the edge, rather than from a physical change in detector response. The fit accurately captures both the transition region, which determines the spatial resolution, and the stable plateau regions far from the edge corresponding to uniform illumination.}
    \label{fig:mtf}
\end{figure}

The modulation transfer function (MTF) for the detector was measured using the slanted knife-edge method, following the procedure described in~\cite{dimova2025measurement}. This method provides a robust estimate of the spatial frequency response of a pixelated detector while mitigating aliasing effects. A custom precision--manufactured aluminium knife-edge plate was fabricated and mounted $0.8~$mm above the sensor surface (Fig.~\ref{fig:mtf}a). The plate provides an edge oriented at a small angle $\theta$ relative to the pixel grid, enabling sub-pixel sampling of the detector response.

A slant angle of $\theta = 5^\circ$ was used. This represents a compromise between achieving sufficient pixel oversampling and ensuring that the measured response approximates the one-dimensional projection of a separable two-dimensional MTF. Larger angles increase oversampling but introduce contributions from diagonal spatial frequencies.

An image was obtained with an overall event rate exceeding $5000$ hits per pixel. This ensured that any pixel-to-pixel variations observed were above statistical (Poisson) noise. A region of interest (ROI) within the illuminated area was selected for analysis (Fig.~\ref{fig:mtf}b). A ROI height of approximately $400$ rows was chosen in order to maximise statistical averaging while avoiding the accumulation of blurring due to imperfections in the knife edge. This was found to provide stable and reproducible results. The recorded data within this region were projected onto an axis perpendicular to the knife edge to obtain the edge spread function (ESF), which represents the detector response to a step-like input signal.

The ESF was constructed using a fixed binning scheme of $n_{\mathrm{bins}} = 8$ per pixel, corresponding to a spatial sampling of $1/8$ pixel. This differs from the ISO~12233 approach~\cite{ISO}, where oversampling is determined by the edge angle. The event-driven nature of Timepix4 results in discrete and non-uniform sampling of the edge, making fixed binning combined with averaging and fitting more robust for this dataset. An ESF obtained at $200~\mathrm{kV}$ is shown in Fig.~\ref{fig:mtf}c.

The ESF was fitted with an error function of the form
\begin{equation}
\mathrm{ESF}(x) = A \, \mathrm{erf}\!\left( \frac{x - \mu}{\sigma} \right) + B,
\end{equation}
where $A$ and $B$ represent scaling and offset, and $\mu$ and $\sigma$ describe the position and width of the transition region. Differentiation of the fitted ESF yields a smooth estimate of the line spread function (LSF), which is subsequently Fourier transformed to obtain the MTF.

The use of a fitted ESF is particularly important for electron detection, where intrinsically noisy data (see Section~\ref{sec:method_gain_factor}) lead to fluctuations in the ESF even after averaging, as shown in Fig.~\ref{fig:mtf}c. Differentiation of such data amplifies these fluctuations, leading to instability in the LSF and MTF at higher spatial frequencies. Fitting the ESF with a smooth analytic function suppresses these stochastic variations and yields a stable and physically meaningful estimate of the detector response.

Prior to Fourier transformation, a Hanning window was applied to the LSF to reduce spectral leakage arising from finite sampling. The measured LSF has finite support, equivalent to multiplication by a rectangular window, which introduces artificial high-frequency components. The Hanning window smoothly suppresses the signal at the boundaries, reducing these artefacts and improving the stability of the calculated MTF.

\FloatBarrier
\subsubsection{Flat-field Data Collection and Processing}
\label{sec:methods_ff_data_collection}

Four flat-field datasets at 200~kV and five datasets at 100~kV were recorded and processed for calculation of the DQE. Each flat-field dataset was recorded with a 50~s exposure and the datasets were saved in HDF5 file format~\mbox{\cite{folk2011overview}} as event lists. An event is a record of the energy deposition caused by an incident electron within a single pixel. Each event contains the pixel index and a timestamp~\cite{llopart2022timepix4} and an event list includes all events recorded within the exposure time on the detector.

The detector was evenly divided into two regions, top and bottom, each with $256 \times 448$ pixels. Sixteen HDF5 files stored the event lists from the top region, and another sixteen files stored event lists from the bottom region.

The exposure time for a single acquisition was limited to approximately 60~s, which corresponds to the time required for a dataset to fill 512~GB of RAM on the control PC. To extend the total exposure time, multiple independent exposures with the same settings were collected, stored, and preprocessed separately before being combined for the NPS calculation.  

The datasets were then processed using the following steps:

\begin{enumerate}
	\item Event lists stored in .h5 files for each dataset were merged and events in the processing area were converted into frames, each containing a target number of events ($N_{tar}$). Here `frames' are defined as 2D image-like representations of the data formed by accumulating events in a pixel array.
	\item Frames at the end of datasets containing fewer than the target number of events ($N_{tar}$) were discarded.
	\item Frames at the same accelerating voltage from multiple datasets were combined and used to calculate the NPS spectrum and zero-frequency DQE.
\end{enumerate}

The target number of events per frame ($N_{tar}$) was set to $5,000,000$ (the summation of each frame generated from the event list is hence also $5,000,000$). In theory, this value should be as low as possible, since a lower value makes the NPS calculation process closer to that handling single-electron events. A lower target number of events ($N_{tar}$) also results in a smoother NPS. However, if the value is too low, the number of frames increases, leading to higher computational time and storage requirements.

As also discussed later pixels along the 448-pixel side of the detector exhibit a higher response than the central area (see Section Raw flat-field data and Fig. S1, S2 in Supplementary Information). We therefore selected a central area of $448 \times 448$ pixels on the detector for the NPS calculation as this gives a more faithful representation of the detector's true performance in practical applications.

The frames converted from the event list can be represented as:

\begin{equation}
	\begin{aligned}
		& c(f, x, y), \quad \text{where:}\\
		& f \in \{1, 2, \dots, N_f\}, \\
		& x \in \{1, 2, \dots, N_x\}, \\
		& y \in \{1, 2, \dots, N_y\},
	\end{aligned}
\end{equation}
where $f$ is the frame index, and $x, y$ are the pixel indices. For the datasets considered in this work, as mentioned above, $N_x = N_y = 448$.

The characteristic properties of $c(f,x,y)$ are described by the parameters $N_e$ and $d_n$. $N_e$ represents the total number of events recorded in the detector area, and $d_n$ represents the mean number of events per pixel per frame, as shown in Equation~\ref{eqa:Ne_def} and Equation~\ref{eqa:dn_def}:

\begin{equation}
    N_e = \sum_{f=1}^{N_f} \sum_{x=1}^{N_x} \sum_{y=1}^{N_y}{c(f,x,y)},
    \label{eqa:Ne_def}
\end{equation}

\begin{equation}
    d_n = \frac{1}{N_f N_x N_y} N_e.
    \label{eqa:dn_def}
\end{equation}

\subsubsection{Gain Factor Calculation}
\label{sec:method_gain_factor}

The average gain factor ($g$) is defined as the ratio of the total number of events recorded in the detector area ($N_e$) to the number of electrons that should have reached the effective area during an exposure time ($N_c$). This is not intrinsic to the detector signal gain, but a result of `charge-sharing'. Charge sharing can arise from the scattering of incident primary electrons across multiple pixels and isotropic lateral diffusion of the generated electron–hole pairs within the sensor, resulting a `cluster' of events. The number of electrons was determined from the effective current ($I_{ef}$) and the exposure time ($t$), given in Equation~\ref{eqa:Nc_def}. The average gain factor $g$ was then calculated from Equation~\ref{eqa:g_cal} as:

\begin{equation}
    N_c = \frac{I_{ef} \cdot t}{e},
    \label{eqa:Nc_def}
\end{equation}

\begin{equation}
	g = \frac{N_e}{N_c}.
	\label{eqa:g_cal}
\end{equation}

In Equation~\ref{eqa:Nc_def}, $e$ represents the charge of an electron. The effective current ($I_{ef}$) was obtained by multiplying the beam current by the ratio of the detector’s effective area to the total illuminating area of the beam. The exposure time ($t$) used in Equation~\ref{eqa:g_cal} accounts only for those frames that contain the target number of events ($N_{tar}$).

The beam current was measured by a Faraday cup and picoammeter. The ratio of the effective illumination area was measured by counting the area (number of pixels) illuminated on the detector with images of the fluorescent screen of known size. Since the beam current was close to the lower detection limit of the picoammeter, the uncertainty of effective beam current ($I_{ef}$), including the uncertainty from ammeter measurement and illumination area, is estimated as approximately 0.5~pA.

\FloatBarrier
\subsubsection{NPS Calculation}
\label{sec:methods_nps_cal}

To calculate the NPS, the first step was to compute a mean flat-field image $c_m(x,y)$ as:

\begin{equation}
	c_{m}(x,y) = \frac{1}{N_f} \cdot \sum_{f=1}^{N_f}{c(f,x,y)}.
	\label{def_c_{m}(x,y)}
\end{equation}
The deviation of the frames from this mean flat-field image $c_m(x,y)$ is:
\begin{equation}
	\Delta(f,x,y) = c(f,x,y) - c_{m}(x,y), \text{ broadcast over }f.
	\label{def_Delta(f,x,y)}
\end{equation}

A 2D Fourier transform with respect to $x$ and $y$ is now performed to calculate the 2D NPS, as in Equation~\ref{cal_NPS_{2D}}. Before performing the Fourier transform, a Hanning window function was applied to $\Delta(f,x,y)$ to reduce the effect of spectral leakage. Following the Fourier transform, the effect of the Hanning window on the amplitude is compensated.

\begin{equation}
	\mathrm{NPS_{2D}}(\omega_x,\omega_y) = \frac{1}{N_f} \cdot \sum_{f=1}^{N_f}{|\mathop{\mathcal{F}}_{x,y} \{ \Delta(f,x,y) \}|^2}.
	\label{cal_NPS_{2D}}
\end{equation}
Since $N_x = N_y = 448$, the sampling of $\mathrm{NPS_{2D}}(\omega_x,\omega_y)$ in $\omega_x$ and $\omega_y$ is the same, i.e. $\omega_x = \omega_y$.

The 1D NPS can be then be computed as a radial average of the 2D NPS, representing a discretised approximation of:

\begin{equation}
	\begin{aligned}
		\mathrm{NPS_{1D}}(\omega_r) &= \frac{1}{2\pi}\int_{0}^{2\pi}{\mathrm{NPS_{2D}}(\omega_r\cos{\theta},\omega_r\sin{\theta})d\theta},\\
		\text{where: } \omega_r&=\sqrt{\omega_x^2 + \omega_y^2}, \: \theta = \arctan \left( \frac{\omega_y}{\omega_x} \right).
	\end{aligned}
\end{equation}

\subsubsection{Estimation of NPS(0) and the NNPS}
\label{sec:method_NPS(0)_est}

Since the mean value of a flat-field image was already subtracted from $\Delta(f,x,y)$ before computing the 2D NPS, the zero-frequency and near zero-frequency components (which correspond to slowly varying brightness in the image) have been removed. As a result, the calculated NPS at zero frequency is underestimated. Therefore, the NPS value of the first and second frequency sampling points were also removed from the NPS curve.

NPS(0) can be computed separately by image binning~\cite{mcmullan2009detective,marcelot2024limitations,paton2023studies}. The frame after binning $B_{Nb}(f, x, y)$ by a binning number $N_b$ can be represented as:

\begin{equation}
	B_{Nb}(f, x, y) = \sum_{a=1}^{N_b} \sum_{b=1}^{N_b} \Delta(f,N_b x + a,N_b y + b).
\end{equation}
For the flat-field datasets used in this work, $B_{Nb}(f, x, y)$ was calculated from $Nb=1$ to $Nb=64$.

The variance of $B_{Nb}(f, x, y)$ is then calculated as $\sigma_{Nb}$:

\begin{equation}
	\sigma_{Nb} = Var[B_{Nb}(f, x, y)] = \frac{1}{N_f N_x N_y} \sum_{f=1}^{N_f} \sum_{x=1}^{N_x} \sum_{y=1}^{N_y} (B_{Nb}(f, x, y) - \overline{B_{Nb}})^2,
\end{equation}
where $\overline{B_{Nb}}$ is the mean value of all pixels in $B_{Nb}$.

As $Nb$ increases, the value of \( \displaystyle \frac{\sigma_{Nb}}{Nb^2} \) converges to a constant which is taken as the value of NPS(0). $\mathrm{NPS_{2D}}$ and $\mathrm{NPS_{1D}}$ are subsequently updated with this value at 0 spatial frequency.

The NNPS can be calculated by normalising the NPS using the calculated value of NPS(0) as:

\begin{equation}
	\mathrm{NNPS}(\omega) = \frac{\mathrm{NPS}(\omega)}{\mathrm{NPS}(0)}.
	\label{eqa:nnps}
\end{equation}

\FloatBarrier
\subsubsection{DQE Calculation}
\label{sec:methods_dqe_cal}

The DQE was calculated from both the MTF and NNPS. The MTF at 100~kV and 200~kV were measured following the method described in~\ref{sec:methods_mtf_meas}. However, since the MTF and NNPS are both normalised at 0 spatial frequency, DQE(0) needs to be calculated separately and subsequently used to normalise the DQE values, as shown in Equations~\ref{eqa:dqe0} and~\ref{eqa:dqe_in_method}.

\begin{equation}
	\mathrm{DQE}(0) = \frac{d_n \cdot g}{\mathrm{NPS}(0)}
	\label{eqa:dqe0}
\end{equation}

In Equation~\ref{eqa:dqe0}, $d_n$ is the mean event recorded per pixel defined in Equation~\ref{eqa:dn_def} and $g$ is the average gain factor defined in Equation~\ref{eqa:g_cal}. Then $\mathrm{DQE}(\omega)$ was finally calculated as~\cite{meyer1998effects,meyer2000characterisation,meyer2000experimental}:

\begin{equation}
	\mathrm{DQE}(\omega)=\mathrm{DQE}(0) \times \frac{\mathrm{MTF}^2(\omega)}{\mathrm{NNPS}(\omega)}.
	\label{eqa:dqe_in_method}
\end{equation}

\subsubsection{Ideal MTF and DQE}
\label{sec:method_ideal_mtf_and_dqe}

The ideal MTF, shown in Fig.~\ref{fig:DQE}, assumes that each pixel of the detector integrates all input events within the detection area without adding noise, meaning that the MTF is determined solely by the uniform sampling of the pixel shape. This ideal MTF ($\mathrm{MTF_{ideal}}$) can be expressed as:

\begin{equation}
	\mathrm{MTF_{ideal}}(\omega) = \left| \frac{\sin(0.5 \pi \omega)}{0.5 \pi \omega} \right|.
	\label{eq:mtf_ideal}
\end{equation}

For an ideal MTF without any consideration of noise, the output signal-to-noise ratio ($\mathrm{SNR_{out}}$) depends only on the MTF applied to the input signal-to-noise ratio ($\mathrm{SNR_{in}}$). In this case, the ideal DQE can be calculated as:

\begin{equation}
	\mathrm{DQE_{ideal}}(\omega) = \frac{\mathrm{SNR_{in}}^2(\omega) \cdot  \mathrm{MTF_{ideal}}^2(\omega)}{\mathrm{SNR_{in}}^2(\omega)}
	\label{eq:dqe_ideal} = \mathrm{MTF_{ideal}}^2(\omega).
\end{equation}

\subsection{Results and Discussion}
\label{sec:dqe_results}

\begin{figure}[htbp]
	\centering
	\includegraphics[width=1.0\textwidth]{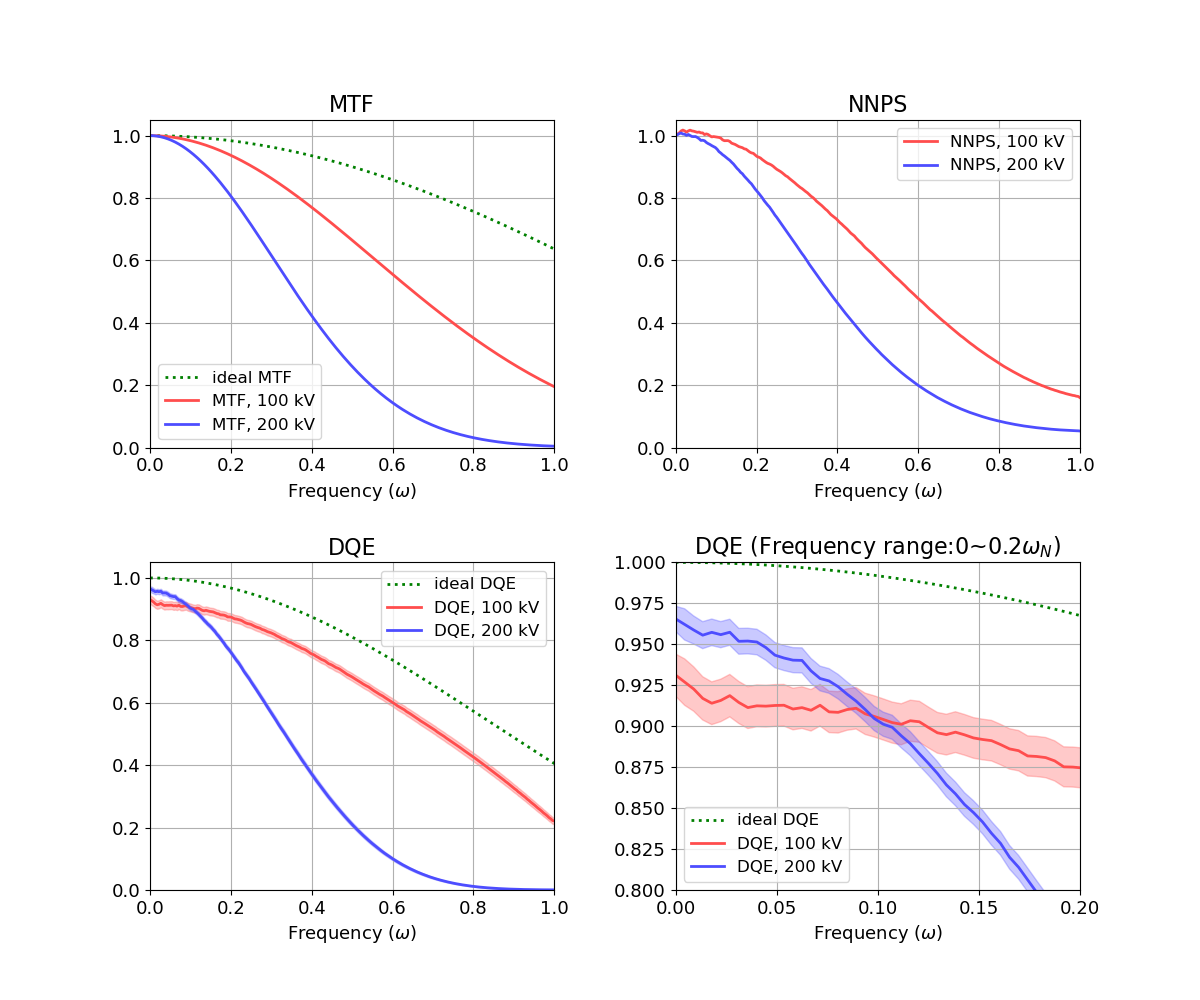}
	\caption{MTF, NNPS and DQE of Timepix4 detector for 100~kV and 200~kV electrons. X-axes are the fraction of spatial frequency to the Nyquist frequency of the detector. The ideal MTF and DQE curves are calculated as described in Section~\ref{sec:method_ideal_mtf_and_dqe}.The semitransparent background of DQE curves indicates estimated uncertainties ($\pm \sigma_{\mathrm{DQE}}$), as discussed in Section~\mbox{\ref{sec:uncertainty_dqe}}. The DQE curves for a frequency range from 0 to 0.2 times the Nyquist frequency ($0.2\omega_N$) are plotted separately (bottom right) showing detailed DQE uncertainties as a semitransparent overlay in this range. The DQE uncertainties $\sigma_{\mathrm{DQE}}$ over the full frequency range from 0 to the Nyquist frequency ($\omega_N$) are plotted in Fig~\mbox{\ref{fig:sigma_mtf_nps_dqe}}.}
	\label{fig:DQE}
\end{figure}

The measured DQE, MTF and NNPS as functions of frequency, from zero to the Nyquist frequency ($\omega_{N}$), are plotted in Fig.~\ref{fig:DQE} for which the DQE, NNPS, average gain factor $g$ (defined by Equation~\ref{eqa:g_cal}) and the zero-frequency DQE were calculated from the flat-field datasets as described in Section~\ref{sec:methods_ff_data_collection}. The MTF was measured as described in Section~\ref{sec:methods_mtf_meas}. The detector settings, including the bias voltage and the per-pixel signal threshold, were kept identical for all measurements in this work.

For a hybrid pixel detector, the MTF, NPS, and DQE are all influenced by its per-pixel threshold. Different threshold settings would yield different calculated results for MTF, NPS, and DQE~\mbox{\cite{paton2021quantifying}}. In this work, the per-pixel threshold for all measurements was set to approximately $3.6~$keV. This is a relatively low threshold compared to the energy of the 100~kV or 200~kV incident electron beam. A low threshold ensures the detector records the electron's incident energy with less losses, resulting in a higher DQE(0). However, a low threshold also leads to more significant charge sharing, degrading the imaging resolution quantified through the MTF. If a higher per-pixel threshold is set, the effect of charge sharing will be reduced. This would result in a lower average gain factor $g$, an improvement in MTF and a flatter NNPS~\mbox{\cite{paton2021quantifying}}. However, since the overall sensitivity of the detector would be reduced, the DQE would decrease, especially at low frequencies. 

The relevant microscope settings and the average gain factor $g$ for the flat-field datasets are listed in Table~\ref{tab:DQE_results}. The calculated values of DQE, MTF, NPS and NNPS at zero frequency, half the Nyquist frequency ($0.5 \, \omega_N$) and the Nyquist frequency ($\omega_N$) are listed in Table~\ref{tab:listed_values}. These values are consistent with previously reported detector characterisations~\cite{mcmullan2007electron,mcmullan2009detective,mir2017characterisation,paton2021quantifying} for the Medipix2 and Medipix3; detectors with the same pixel pitch and sensor type as the Timepix4 assembly used in this work.

\begin{table}[htbp]
	\centering
	\caption{Microscope settings and average gain factor ($g$) for flat-field datasets}
	\begin{tabular}{|c|c|c|}
		\hline
		Accelerating Voltage & 100~kV & 200~kV \\
		\hline
		Effective Beam current on detector, $I_{ef}$ & $14.4 \pm 0.5$~pA & $13.0 \pm 0.5$~pA \\
		\hline
		Exposure time, $t$ & 248.7~sec & 198.0~sec \\
		\hline
		Number of events on detector, $N_e$ & 56,155,000,000 & 77,595,000,000 \\
		\hline
		Number of electrons by beam current, $N_c$, & 22,311,450,649 & 16,035,553,305 \\
        \hline
		Average gain factor, $g$ & $2.517 \pm 0.087$ & $4.840 \pm 0.186$ \\
		\hline
	\end{tabular}
	\label{tab:DQE_results}
	\\[5pt]
\end{table}

\begin{table}[htbp]
    \centering
    \caption{DQE, MTF, NNPS and NPS measured at zero frequency, half the Nyquist frequency ($0.5 \, \omega_N$) and the Nyquist frequency ($\omega_N$).}
    \begin{tabular}{|c|c|c|c|}
    \hline
    Frequency, $\omega$ & $0$ & $0.5 \, \omega_N $ & $\omega_N$ \\
    \hline
    DQE, 200~kV & $0.965 \pm 0.008$ & $0.210 \pm 0.007$ & $(3.81 \pm 0.51) \times 10 ^ {-4}$ \\
    \hline
    DQE, 100~kV & $0.931 \pm 0.013$ & $0.682 \pm 0.012$ & $0.221 \pm 0.011$  \\
    \hline
    DQE, ideal & $1.000$ & $0.811$ & $0.405$ \\
    \hline
    MTF, 200~kV & 1* & $0.260 \pm 0.004$ & $0.0046 \pm 0.0003$ \\
    \hline
    MTF, 100~kV & 1* & $0.665 \pm 0.004$ & $0.195 \pm 0.004$ \\
    \hline
    MTF, ideal & 1* & $0.900$ & $0.637$ \\
    \hline
    NNPS, 200~kV & 1* & $0.3114 \pm 0.0003$ & $0.0533 \pm 0.0001$ \\
    \hline
    NNPS, 100~kV & 1* & $0.6029 \pm 0.0008$ & $0.1602 \pm 0.0002$ \\
    \hline
    NPS, 200~kV & $124.9 \pm 0.1$ & $38.88 \pm 0.02$ & $6.654 \pm 0.003$ \\
    \hline
    NPS, 100~kV & $67.35 \pm 0.07$ & $40.61 \pm 0.03$ & $10.791 \pm 0.006$ \\
    \hline
    \end{tabular}\\
    [5pt]
    *: Normalised value
    \label{tab:listed_values}
\end{table}

The average gain factor $g$ of the Timepix4 detector was measured as 2.517 at 100~kV and 4.840 at 200~kV using the threshold and reverse-bias settings in Section~\ref{sec:detector_system}. This observed difference in the average gain factor $g$ originates from charge sharing in which at higher electron energies, electrons deposit energy over a larger interaction volume and interact with more adjacent pixels in a cluster. Combined with a finite threshold, this leads to an increase in the number of recorded events and results in a larger measured average gain factor $g$. Further, this increased interaction volume also results in a broader point spread function for a single electron event and consequently reduces the MTF at 200~kV compared to 100~kV, as shown in Fig.~\ref{fig:DQE}.

For this reason, the experimental MTF can be enhanced and brought closer to the ideal MTF by applying a clustering algorithm~\cite{van2020sub,dimova2025measurement}. However, in this work, all measurements are based on the raw data from the Timepix4 detector without any clustering applied.

The NNPS, as shown in Fig.~\ref{fig:DQE}, is also affected by charge sharing which introduces spatial correlations between neighbouring pixels and effectively acts as a low-pass filter, suppressing high-frequency noise components. As a result, the NNPS measured at 200~kV shows a faster decline with increasing spatial frequency compared to that at 100~kV, for unchanged detector parameters.

With charge sharing affecting both the MTF and NNPS, the DQE at 100~kV has a value of 0.221 at the Nyquist frequency, where the ideal DQE (described in Section~\ref{sec:method_ideal_mtf_and_dqe}) is 0.405. However at 200~kV, the DQE at the Nyquist frequency is close to zero ($3.81 \times 10 ^{-4}$) mainly because of the reduced MTF. At half the Nyquist frequency, as shown in Table~\ref{tab:listed_values}, the DQE at 100~kV is 0.682, whereas at 200~kV it is 0.210, which is approximately one quarter of the ideal value of 0.811.

The semitransparent background in Fig.~\ref{fig:DQE} and the values indicated by `$\pm$' in Table~\ref{tab:DQE_results} and Table~\ref{tab:listed_values} represent the estimated uncertainties. The estimated uncertainties from zero to the Nyquist frequency in the DQE at 100~kV is less than 0.015 and at 200~kV is less than 0.010. A detailed description of uncertainty estimation is given in Section~\ref{sec:uncertainty_estimation}. 

In the zero-frequency and low-frequency regions (below 1/10 of Nyquist frequency), the DQE at 200~kV is slightly higher than that at 100~kV. At zero frequency, the measured DQE(0) at 200~kV is 0.965, while at 100~kV it is only 0.931, as shown in Table~\ref{tab:listed_values} and Fig~\mbox{\ref{fig:DQE}}. The difference in DQE(0) at 100 kV and 200 kV arises due to the skewness in the cluster size distribution. The 100 kV cluster size distribution is more positively skewed than at 200 kV, as shown in Fig. S3 in the Supplementary Information.

As previously noted, clustering algorithms designed to compensate for charge sharing effects have been shown to improve the MTF of Timepix4 detectors~\cite{van2020sub,dimova2025measurement}. These are especially powerful for data recorded with a low signal threshold, where large cluster sizes provide information about the charge deposition in the event. As the NNPS is also influenced by charge sharing, it is expected that such clustering algorithms would also flatten the NPS and, consequently, alter the DQE.

While such approaches can improve performance by mitigating charge sharing, they introduce additional processing-dependent effects that fall outside the scope of this work, which focuses on the intrinsic performance of the detector in raw event-driven mode.

\subsection{Uncertainty Estimation}
\label{sec:uncertainty_estimation}
\subsubsection{Uncertainty in the DQE}
\label{sec:uncertainty_dqe}

The uncertainty in the DQE at each spatial frequency, $\sigma_{\mathrm{DQE}}(\omega)$, was obtained using a standard first-order propagation of independent uncertainties in all contributing quantities. In Equation~\ref{eqa:dqe_error}, $\sigma_{\mathrm{DQE}}$ is represented by uncertainties of the average gain factor $g$, the mean number of events per pixel per frame $d_n$, the $\mathrm{NPS}(\omega)$, and the $\mathrm{MTF}(\omega)$:

\begin{equation}
    \sigma^2_{\mathrm{DQE}}(\omega) =
    \left(\frac{\partial \mathrm{DQE}}{\partial g}\right)^2 \sigma_g^2
    +
    \left( \frac{\partial \mathrm{DQE}}{\partial d_n}\right)^2 \sigma_{d_n}^2
    +
    \left(\frac{\partial \mathrm{DQE}}{\partial \mathrm{MTF}}\right)^2 \sigma_{\mathrm{MTF}}^2(\omega)
    +
    \left(\frac{\partial \mathrm{DQE}}{\partial \mathrm{NPS}}\right)^2 \sigma_{\mathrm{NPS}}^2(\omega) .
    	\label{eqa:dqe_error}
\end{equation}

Based on Equation~\ref{eqa:dqe_error}, Equation~\ref{eqa:dqe0} and Equation~\ref{eqa:dqe_in_method}, $\sigma_{\mathrm{DQE}}$ can be represented as:

\begin{equation}
    \sigma_{\mathrm{DQE}}(\omega) = k_{\mathrm{DQE}}(\omega) \, \mathrm{DQE}(\omega)
    \label{eq:sigma_dqe_by_k}
\end{equation}
with $k_{DQE}$ expressed as:

\begin{equation}
    k_{\mathrm{DQE}}(\omega) = \sqrt{\left( \frac{\sigma_g}{g} \right)^2 
    + \left( \frac{\sigma_{d_n}}{d_n} \right)^2
    + \left( \frac{2 \, \sigma_{\mathrm{MTF}}(\omega)}{\mathrm{MTF}(\omega)} \right)^2
    + \left( \frac{\sigma_{\mathrm{NPS}}(\omega)}{\mathrm{NPS}(\omega)} \right)^2
    }.
    \label{eq:k_dqe}
\end{equation}

\begin{figure}[htbp]
	\centering
	\includegraphics[width=1.0\textwidth]{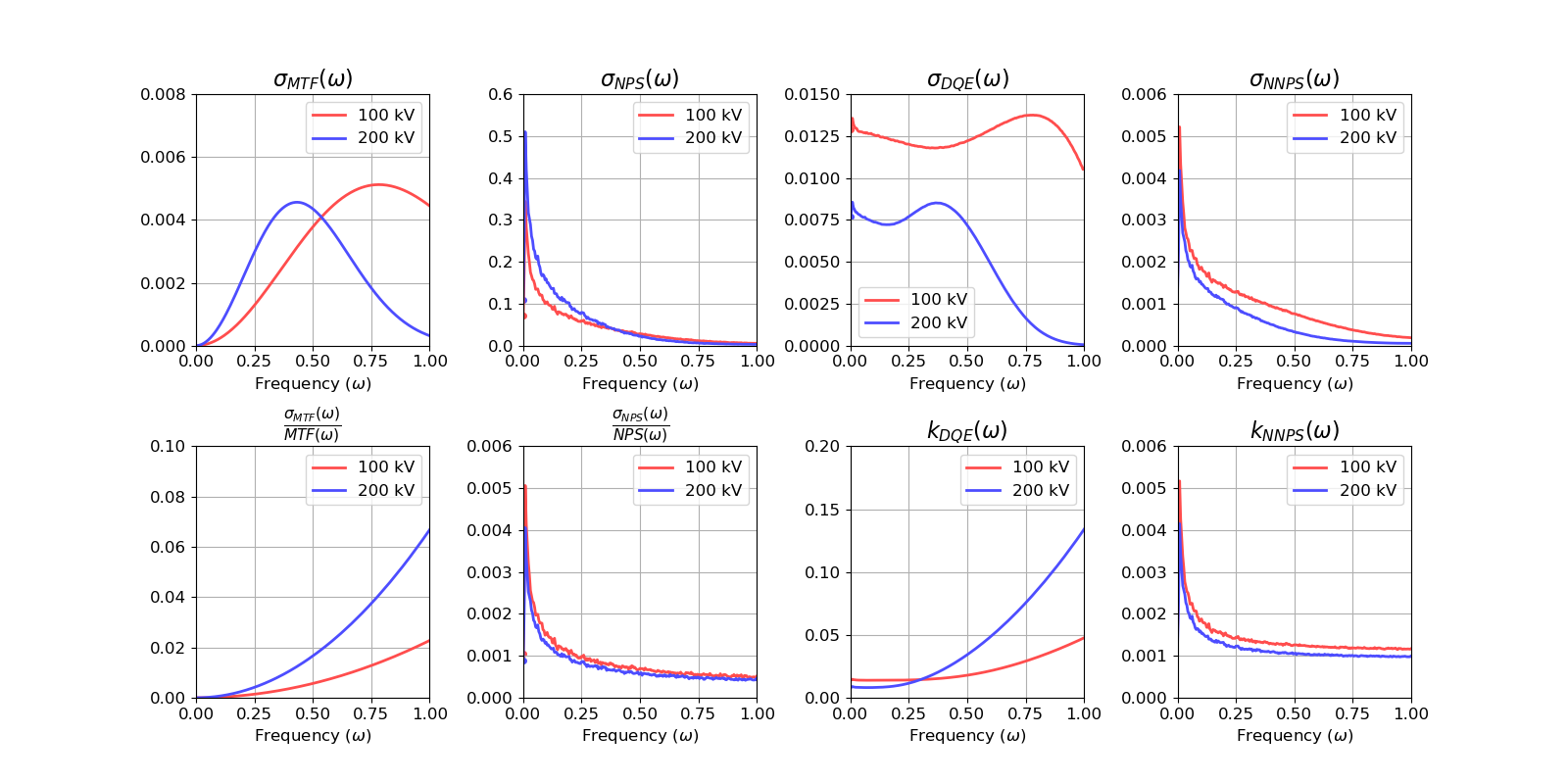}
	\caption{$\sigma_{\mathrm{MTF}}$, $\sigma_{\mathrm{NPS}}$, $\sigma_{\mathrm{DQE}}$, $\sigma_{\mathrm{NNPS}}$, $\frac{\sigma_{\mathrm{MTF}}}{\mathrm{MTF}}$, $\frac{\sigma_{\mathrm{NPS}}}{\mathrm{NPS}}$, $k_{\mathrm{DQE}}$ and $k_{\mathrm{NNPS}}$ for 100~kV and 200~kV. X-axes are the ratio of spatial frequency to the Nyquist sampling frequency of the detector. $\sigma_{\mathrm{NPS}}(0)$ and $\frac{\sigma_{\mathrm{NPS}}(0)}{\mathrm{NPS}(0)}$ are marked as dots on y-axis.}
	\label{fig:sigma_mtf_nps_dqe}
\end{figure}

The uncertainty in the MTF ($\sigma_{\mathrm{MTF}}$) is determined through a Monte Carlo propagation of the edge-spread function (ESF) fit parameters; after fitting the ESF, random parameter vectors were sampled from the multivariate normal distribution defined by the best-fit parameters and their covariance. For each sample, the ESF, a line-spread function, and the resulting MTF were then recomputed. The spread of this ensemble at each spatial frequency provides the MTF standard deviation $\sigma_{\mathrm{MTF}}$. To verify that the MTF uncertainty distribution is sufficiently symmetric for linear error propagation, the standard deviation was compared against the non-parametric 16th–84th percentile interval of the Monte-Carlo ensemble. The percentile-based half-width closely matches $\sigma_{\mathrm{MTF}}$ with a mean relative difference of $\ll 1$ across all spatial frequencies. Additionally, the RMS difference between the upper and lower percentile deviations normalised by the standard deviation is $< 0.1$, indicating negligible skewness. Hence, the standard deviation was used as the MTF uncertainty term in the DQE error propagation and $\sigma_{\mathrm{MTF}}$ at 100~kV and 200~kV are plotted in Fig.~\ref{fig:sigma_mtf_nps_dqe}.

The uncertainty in $g$ was treated as a frequency independent scalar quantity. The value of $g$ was calculated using the total number of events in all frames ($N_e$, defined in Equation~\ref{eqa:Ne_def}) and the estimated number of electrons incident in the detector area during the exposure time ($N_c$, defined in Equation~\ref{eqa:Nc_def}). The quantity $N_e$ is a fixed value determined by the conversion of the event list into frames and by $N_{tar}$ (see Section~\ref{sec:methods_ff_data_collection}~\nameref{sec:methods_ff_data_collection}). Consequently, the uncertainty in $g$ is solely determined by the uncertainty in the effective beam current measurement $I_{ef}$. The $I_{ef}$ (13 to 14~pA as shown in Table~\ref{tab:DQE_results}) used here is close to the lower detection limit of the Faraday cup and the ammeter used and the uncertainty in $I_{ef}$ is estimated to be 0.5~pA based on the precision of the ammeter. Finally, $\sigma_g/g$ was calculated as $0.5/I_{ef}$, approximately 0.035 at 100~kV and 0.038 at 200~kV.

The uncertainty $\sigma_g$ represents the uncertainty in the mean gain factor and is distinct from the intrinsic variance in the detector response per incident electron. The latter contributes to the noise power spectrum and is therefore accounted for in $\mathrm{NPS}(0)$, rather than in $\sigma_g$.

The uncertainty in the NPS was treated as per-frequency standard deviations taken from repeated measurements. The Bootstrap method was used to estimate the standard deviation of NPS. From all $N_f$ flat-field frames, $N_f$ frames were randomly resampled with replacement, and the frames selected in each sampling were treated as an independent dataset for the NPS calculation. This procedure was repeated $n_{bs}$ times. Subsequently, the standard deviation at each spatial frequency was calculated from the $n_{bs}$ independently computed NPS results. This standard deviation is then taken to be $\sigma_{\mathrm{NPS}}$, plotted in Fig.~\ref{fig:sigma_mtf_nps_dqe} at 100~kV and 200~kV. The value of $n_{bs}$ was set to 500 for the calculations in this work.

In Fig.~\ref{fig:sigma_mtf_nps_dqe}, $\sigma_{\mathrm{NPS}}(0)$ at both 100~kV and 200~kV are marked as dots on y-axis. This is because, in the Bootstrap method, $\sigma_{\mathrm{NPS}}(0)$ were separately calculated by binning as described in Section~\ref{sec:method_NPS(0)_est}, which results in a discontinuity at zero frequency.

The uncertainty of $\sigma_{d_n}$, the mean number of events per pixel per frame, was calculated from the standard error of the mean (SEM) of all flat-field frames $c(f,x,y)$, as the SEM indicates by how much the $d_n$, the mean value of $c(f,x,y)$, is expected to vary from the true mean value under the current sample size. The $\sigma_{d_n}$ was calculated as:

\begin{equation}
    \sigma_{d_n} = \frac{s_c}{\sqrt{N_f N_x N_y}},
    \label{eq:sigma_dn}
\end{equation}
where $s_c$ is the standard deviation of $c(f,x,y)$ with $s_c$, $d_n$, $\sigma_{d_n}$ and related parameters are listed in Table~\ref{tab:frame_num_and_mean}. This gives values of $\sigma_{d_n}$ as $1.050 \times 10 ^{-4}$ at 100~kV and $8.924 \times 10 ^ {-5}$ at 200~kV.

By substituting all the results for $\sigma_g$, $\sigma_{d_n}$, $\sigma_{\mathrm{MTF}}$ and $\sigma_{\mathrm{NPS}}$ into Equation~\ref{eq:sigma_dqe_by_k} and Equation~\ref{eq:k_dqe}, $k_{\mathrm{DQE}}$ and $\sigma_{\mathrm{DQE}}$ can be then calculated and are plotted in Fig.~\ref{fig:sigma_mtf_nps_dqe}.

\begin{table}[htbp]
    \centering
    \caption{$d_n$, $N_{tar}$ for DQE calculations}
    \begin{tabular}{|c|c|c|}
    \hline
    Accelerating Voltage& 100 kV & 200 kV \\
    \hline
    Target number of events per frame, $N_{tar}$ & 5,000,000 & 5,000,000 \\
    \hline
    Number of frames, $N_f$ & 11,231 & 15,519 \\
    \hline
    Average number of events per pixel per frame, $d_n$ & 24.91 & 24.91 \\
    \hline
    Standard deviation of all pixels on all frames, $s_c$ & 4.985 & 4.980 \\
    \hline
    SEM of all pixels on all frames, $\sigma_{d_n}$ & $1.050 \times 10 ^{-4}$ & $8.924 \times 10 ^ {-5}$ \\
    \hline
    \end{tabular}
    \label{tab:frame_num_and_mean}
    \\[5pt]
\end{table}

The calculated $\sigma_{\mathrm{DQE}}$ at 100~kV and 200~kV are shown in Table~\ref{tab:listed_values} and in Fig.~\ref{fig:DQE} as the semitransparent background of DQE curves. In general, the estimated uncertainty of measured DQE is less than 0.015 at 100~kV and 0.010 at 200~kV from zero frequency to the Nyquist frequency. The calculated $\sigma_{\mathrm{MTF}}$, $\sigma_{\mathrm{NPS}}$ and $\sigma_{\mathrm{NNPS}}$ at half the Nyquist frequency ($0.5 \, \omega_N$) and the Nyquist frequency ($\omega_N$) are listed in Table~\ref{tab:listed_values}.

\FloatBarrier
\subsubsection{Uncertainty in the NNPS}
\label{sec:uncertainty_nnps}

The uncertainty of NNPS for $\omega>0$ is derived from Equation~\ref{eqa:nnps} using error propagation. Since $\mathrm{NPS}(0)$ is calculated independently of $\mathrm{NPS}(\omega)$, their uncertainties, $\sigma_{\mathrm{NPS}(0)}$ and $\sigma_{\mathrm{NPS}(\omega)}$, are independent. Therefore, the error propagation can be expressed as in Equation~\ref{eqa:sigma_nnps_1}.

\begin{equation}
    \sigma_{\mathrm{NNPS}} ^{2} (\omega) = \left( \frac{\partial \mathrm{NNPS}(\omega)}{\partial \mathrm{NPS}(0)} \right)^2 \sigma_{\mathrm{NPS}(0)}^2 
    + \left( \frac{\partial \mathrm{NNPS}(\omega)}{\partial \mathrm{NPS}(\omega)} \right)^2 \sigma_{\mathrm{NPS}(\omega)}^2 \;, \mathrm{when} \;\omega>0 
    \label{eqa:sigma_nnps_1}
\end{equation}

The uncertainty $\sigma_\mathrm{NNPS}(\omega)$ can be calculated by multiplying $\mathrm{NNPS}(\omega)$ and $k_\mathrm{NNPS}(\omega)$, as shown in Equation~\ref{eqa:sigma_nnps_k}. Here, $k_\mathrm{NNPS}(\omega)$ is expressed in terms of $\mathrm{NPS}(\omega)$, $\mathrm{NPS}(0)$, $\sigma_{\mathrm{NPS}}(\omega)$, and $\sigma_{\mathrm{NPS}}(0)$, as given in Equation~\ref{eqa:sigma_nnps_3}.

\begin{equation}
    \sigma_{\mathrm{NNPS}}(\omega) = k_\mathrm{NNPS}(\omega) \, \mathrm{NNPS}(\omega)
    \label{eqa:sigma_nnps_k}
\end{equation}

\begin{equation}
    k_\mathrm{NNPS}(\omega) = \sqrt{\frac{\sigma_{\mathrm{NPS}}^2(0)}{\mathrm{NPS}^2(0)} + \frac{\sigma_{\mathrm{NPS}}^2(\omega)}{\mathrm{NPS}^2(\omega)}}
    \label{eqa:sigma_nnps_3}
\end{equation}

The plots of $\sigma_{\mathrm{NNPS}}(\omega)$ and $k_{\mathrm{NNPS}}(\omega)$ are shown in Fig.~\ref{fig:sigma_mtf_nps_dqe}. The values of $\sigma_{\mathrm{NNPS}}$ at $0.5 \,\omega_N$ and $\omega_N$ are listed in Table~\ref{tab:listed_values}. The $\sigma_{\mathrm{NNPS}}$ from zero to the Nyquist frequency is below 0.006, which is negligible compared to the calculated NNPS. 

\subsubsection{Other minor factors}
\label{sec:uncertainty_other_minor_factors}

In addition to the uncertainties considered above, the accuracy of the DQE measurement may also be affected by the following factors:

\begin{enumerate}

	\item The total events in each dataset decreases with the duration of the data collection. During the approximately two-hour acquisition period used, multiple datasets showed a trend in which the total number of events in the final dataset acquired was approximately 0.5\% lower than in the initial dataset. This was observed at both 100~kV and 200~kV and introduces an error in the calculation of the average gain factor ($g$). We suggest that this effect is due to a reduction in the emission current of the cold-FEG over the duration of the acquisition.

	\item As noted earlier the detector response is not uniform. The outer area, especially along the two 448-pixel edges which are closest to the readout periphery, showed a slightly higher response compared to the central area which is primarily used in the measurements described. This results in an overestimation of the low-frequency component of the NPS. Therefore, only the central $448 \times 448$ pixels were used to reduce this skewing.

	\item There are 45 dead pixels visible at 200 kV and 46 dead pixels at 100 kV where the value of a dead pixel is always 0. These dead pixels lead to a lower total number of events ($N_e$) and a slight underestimation of average gain factor ($g$). However, dead pixels account for only about 0.02\% of all pixels within the $448 \times 448$ detector area, and thus the resulting underestimation of $g$ due to dead pixels is also expected to be of the order of 0.02\%. This error is much smaller than the uncertainty in $g$ arising from the current measurement ($I_{ef}$) and was therefore neglected by the analysis described. In addition, since the NPS calculation is based on the deviation of the frames from the mean flat-field image (see Equation~\ref{def_Delta(f,x,y)}), dead pixels with zero values do not affect the NPS calculation. 

    \item The temperature of the detector chip increased from an initial value of 20 degree Celsius and stabilised at approximately 70 degree Celsius within two hours after the detector was initially powered on. The influence of temperature on the detector response was not measured, as no observable change in response was recorded during this period. Theoretically, the temperature affects only the mobility of charge carriers, rather than the number of electron-hole pairs generated. It has been reported that there is only a small temperature dependence in detector response parameters within typical operating ranges~\cite{rose_2024_acwjn-22672}, suggesting that temperature effects are unlikely to be significant.
\end{enumerate}

The amount of data collected in this work is sufficient to support the above statistical analysis. The number of events collected ranges from $5.6 \times 10^{10}$ (100~kV) to $7.7 \times 10^{10}$ (200~kV), as shown in Table~\ref{tab:DQE_results}, corresponding to an average of $2.8 \times 10^{5}$ (100~kV) and $3.8 \times 10^{5}$ (200~kV) events per pixel, respectively. 

\FloatBarrier
\section{Parallel-Beam Diffraction}

\subsection{Experiment and results}

\begin{figure}[!htbp]
	\centering
	\begin{subfigure}{0.6\textwidth}
		\centering
		\begin{overpic}[width=\linewidth]{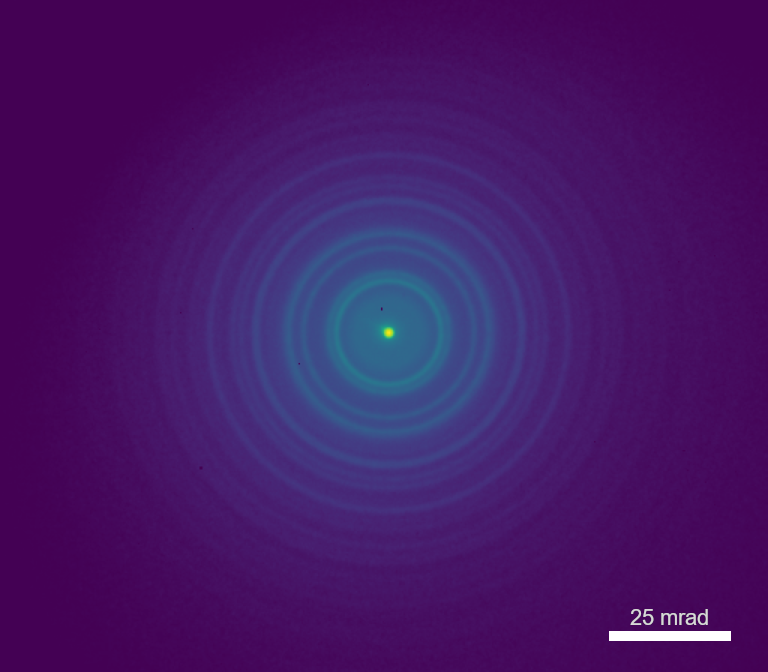}
			\put(-10,85){(a)}
		\end{overpic}
		\label{fig:TP4_rings_log_scale}
	\end{subfigure}
	\begin{subfigure}{0.8\textwidth}
		\begin{overpic}[width=\linewidth]{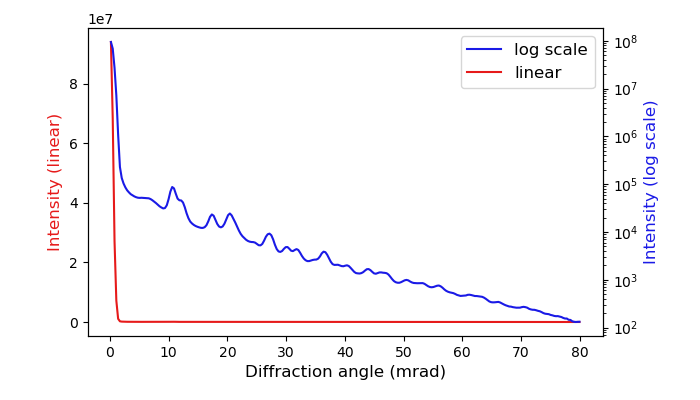}
			\put(3,52){(b)}
		\end{overpic}
		\label{fig:lp_0-80mra}
	\end{subfigure}
	\caption{Parallel-Beam Diffraction pattern from a polycrystalline gold sample recorded on a Timepix4 detector at 200 kV and corresponding average radial line profiles. (a) Diffraction pattern recorded on a Timepix4 detector. Colormap is log-scale. (b) Average radial line profile of the diffraction pattern in (a), plotted on both linear (red line) and log scales (blue line).}
	\label{fig:diff_rings_and_lp}
\end{figure}

As an initial demonstration of Timepix4 detector for electron diffraction, a polycrystalline diffraction pattern of a standard gold polycrystalline nanoparticle sample was recorded without a direct beam stop. A typical diffraction pattern is shown in Fig.~\ref{fig:diff_rings_and_lp}(a) on a log-scale colormap. 

The same microscope (JEOL CryoARM Z300FSC) and detector setting as for flat-field data were used to record diffraction data. Each diffraction pattern was recorded at 200~kV with a 20~s exposure giving a total number of electrons at the sample of approximately $1.7 \times 10 ^ {10}$ and a fluence at the sample of approximately $3.2 \times 10 ^{-2}$~e\textsuperscript{-}/nm\textsuperscript{2}. However, since a selected area aperture and energy filter were used for data recording, the number of electrons on the detector was lower than on the sample. The number of events recorded was 184,227,225 with a corresponding number of electrons of $3.8 \times 10 ^7$, estimated by the average gain factor $g$ at 200~kV (4.840). Therefore, only about 0.2\% of the electrons that are incident on the sample are recorded on the detector.

Fig.~\ref{fig:diff_rings_and_lp}(b) shows averaged radial line profiles from the diffraction pattern on a log-scale (blue) and a linear-scale (red) plotted on the same X-axis. 

\begin{figure}[htbp]
	\centering
	\includegraphics[width=0.8\textwidth]{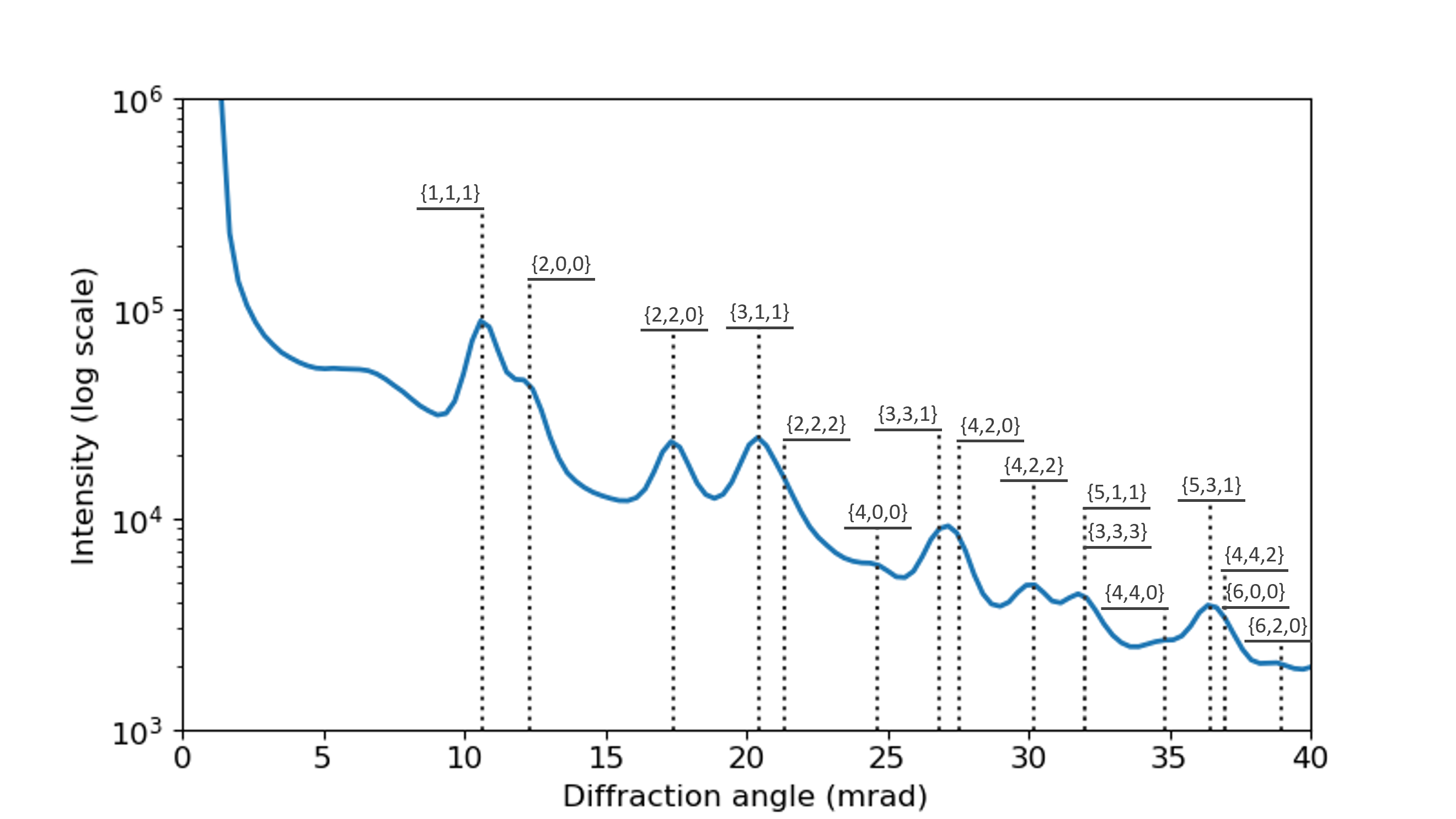}
	\caption{Average radial line profile of the diffraction pattern (0 to 40~mrad, corresponding to a limiting d-spacing of 0.06275~nm) and indexed gold reflections. The line profile is plotted on log scales. Indexed reflections are indicated by black dotted lines.}
	\label{fig:lp_0-40mrad_with_refl}
\end{figure}

The 4 lowest angle reflections (corresponding to \{111\},\{200\},\{220\},\{311\}) were used for calibration of the pixel size at the diffraction plane, giving a pixel size of 0.307~mrad/pixel. Reflections below 40~mrad are indexed and labelled in Fig.~\ref{fig:lp_0-40mrad_with_refl}.

\FloatBarrier
\subsection{Discussion}

\begin{figure}[htbp]
	\centering
	\includegraphics[width=1.0\textwidth]{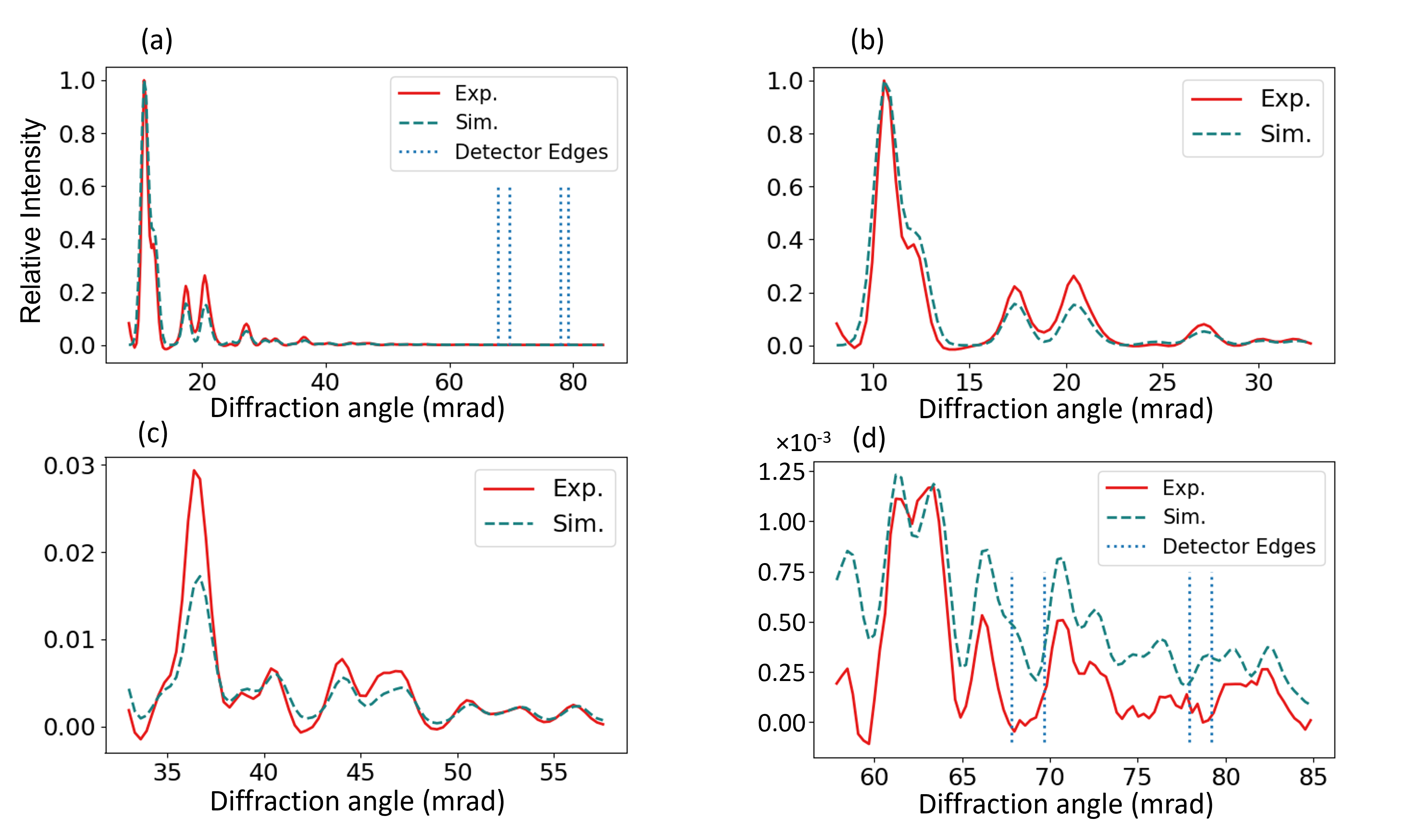}
	\caption{Experimental average radial line profile (red solid line, background removed) and simulated diffraction intensity distribution (green dash line). The blue dotted lines indicate the detector edges. (a) line profiles from 8 to 85~mrad (corresponding to d-spacings from 0.3138 to 0.0295~nm). (b), (c) and (d) line profiles in the range of 8 to 35~mrad (corresponding to d-spacings from 0.3138 to 0.0717~nm), 30 to 60~mrad (corresponding to d-spacing from 0.0837 to 0.0418~nm) and 55 to 85~mrad (corresponding to d-spacing from 0.0456 to 0.0295~nm), respectively with different y-axis scales applied.}
	\label{fig:lp_exp_vs_sim}
\end{figure}

Diffraction at angles higher than 40~mrad (corresponding to a d-spacing of approximately 0.06275~nm), average radial line profiles of the diffraction patterns for different angular ranges are shown in Fig.~\ref{fig:lp_exp_vs_sim}. The average radial line profile was processed using an Asymmetric Least Squares (ALS) method~\cite{eilers2005baseline} (with parameters $\lambda=100$, $p=0.01$ and 100 iteration, mirror-padded at both ends) to remove the background. The intensity of the line profile was normalised to the intensity of the reflection \{111\}. In Fig.~\ref{fig:lp_exp_vs_sim}(a) the average line profile between angles 8 to 85~mrad (corresponding to d-spacings from approximately 0.3138 to 0.0295~nm) is plotted.

In Fig.~\ref{fig:lp_exp_vs_sim}(d), the average radial line profile of the diffraction pattern shows variations in the relative intensity from 0 to approximately $1.2 \times 10 ^ {-3}$. To determine whether this variation is due to overlap of multiple diffraction peaks or simply noise, a simulated intensity distribution was calculated shown as the green dashed curve in Fig.~\ref{fig:lp_exp_vs_sim}.

The simulated diffraction intensity distribution was calculated using the ReciPro software~\cite{seto2022recipro} using the FCC gold structure reported in~\cite{suh1988high} and electron scattering factors reported in~\cite{peng1996robust}. An electron wavelength of 2.51~pm at 200~kV was used for calculations. Individual reflections were modelled using Gaussian functions. A standard deviation for the Gaussian function was set to 0.6~mrad for all reflections and the height of Gaussian functions used was proportional to the relative intensities of the reflections normalised by the \{111\} intensity.

The simulated diffraction distribution does not allow quantitative comparison, since many factors that affect the intensity distribution were not considered including, for example, crystal grain size and Debye-Waller factors. However, the simulated diffraction intensity distribution exhibits a trend similar to that of the experimentally obtained radial average line profile in the angle range of 60 to 75~mrad (corresponding to d-spacings from 0.0418 to 0.0335~nm), as shown in Fig.~\ref{fig:lp_exp_vs_sim}(d). This demonstrates that variations in the diffraction data up to at least 75~mrad (corresponding to a d-spacing of 0.0355~nm) on the experimental line profile is not noise, but is formed from an overlap of diffracted intensities.

To quantitatively demonstrate the capability to resolve weak signals from diffraction patterns and average radial line profiles, the effective intensity span ($C_e$) of the line profile was calculated. The effective intensity span ($C_e$) was evaluated from the radial average line profile of the diffraction pattern, which defines maximum intensity ($I_{max}$) and the lowest resolvable intensity ($I_{min,r}$) in reciprocal space after angular integration. The value of $C_e$ was then calculated using Equation~\ref{eq:C_e_def}. The intensity in this analysis indicates the total number of events recorded on a pixel, without clustering and other post-processing as:

\begin{equation}
    C_e = \frac{I_{max}}{I_{min,r}}.
    \label{eq:C_e_def}
\end{equation}

As discussed above, in Fig.~\ref{fig:lp_exp_vs_sim}(d), the radial average line profile shows diffraction information up to 75~mrad (corresponding to a d-spacing of 0.0355~nm). Therefore, the resolvable minimum intensity ($I_{min,r}$) for the average line profile is estimated as 186, the mean intensity at 75~mrad diffraction angle. The maximum intensity ($I_{max}$) is $1.143 \times 10 ^{7}$ at the centre beam. The effective intensity span ($C_e$) for the average line profile is then calculated as $6.15 \times 10^4$.

However, the $C_e$ using this calculation is underestimated. Approximately 27\% of the events recorded at the centre beam pixel are recorded with a `pileup' flag which means that this event has an abnormally high signal triggered in detector. This usually means multiple electrons arriving at a pixel within a short period of time such that the detector cannot separate them. As an estimation, the maximum intensity ($I_{max}$) could be at least 27\% higher if every single electron is recorded separately. However, `pileup' events do not influence the resolvable minimum intensity ($I_{min,r}$), since the number of `pileup' events at a 75~mrad diffraction angle is 0.

This initial application shows the capability of the Timepix4 detector to resolve weak signals in electon diffraction patterns, achieving an effective intensity range of at least $6.15 \times 10^4$. When combined with the detector’s high zero-frequency detective quantum efficiency (DQE) at greater than 0.9, the calculated effective intensity range shows that the detector has potential for detecting weak diffracted intensities.

It should be noted, however, that the effective intensity range reported in this demonstration is calculated from radial average line profile, rather than being calculated directly from individual pixels. This radial averaging reduces noise in pixels receiving a low number of electron events and hence the reported value should not be directly interpreted as the effective intensity range in the raw diffraction patterns. 

\FloatBarrier
\section{Conclusion} 

In this work, the NNPS and DQE of a Timepix4 detector operated in a transmission electron microscope at 100~kV and 200~kV have been measured without clustering. The zero-frequency DQE values were measured to be 0.931 at 100~kV and 0.965 at 200~kV. At the Nyquist frequency, the DQE at 100~kV is 0.221, whereas at 200~kV it decreases to $3.81 \times 10^{-4}$. Uncertainties of the DQE measurements are estimated to be less than 0.015 at 100~kV and less than 0.010 at 200~kV.

An example diffraction image recorded by the Timepix4 detector is also shown. The radial average line profile of the diffraction pattern shows diffracted information extending to at least 75~mrad (corresponding to a d-spacing of 0.0355~nm), with an effective intensity range in a calculated radial average of $6.15 \times 10^{4}$. Given its fast readout speed, the Timepix4 detector is ideally suited for high-resolution, low-fluence electron diffraction.

\section{Contributions}

\noindent Zhiyuan Ding: Writing the manuscript; Review and editing; MTF, Flat-field and diffraction data recording; Flat-field and diffraction data analysis; Error estimation; Visualization.

\noindent Nina Dimova: Review and editing; MTF, Flat-field data recording; MTF data analysis; Error estimation. 

\noindent Jonathan S. Barnard: Review and editing; MTF, Flat-field and diffraction data recording.

\noindent Giulio Crevatin: Review and editing; MTF and Flat-field data recording; Software support.

\noindent Liam O'Ryan: Review and editing.

\noindent Richard Plackett: Review and editing.

\noindent Daniela Bortoletto: Review and editing.

\noindent Angus I. Kirkland: Review and editing; Overall project supervision.

\noindent Marcus Gallagher-Jones: Review and editing; MTF, Flat-field and diffraction data recording.

\appendix
\section*{Supplementary Information}
\renewcommand{\thesection}{S\arabic{section}}

\section{Raw Flat-field image}
\label{secS:raw_flat_field_image}

\begin{figure}[htbp]
	\centering
	\includegraphics[width=0.8\textwidth]{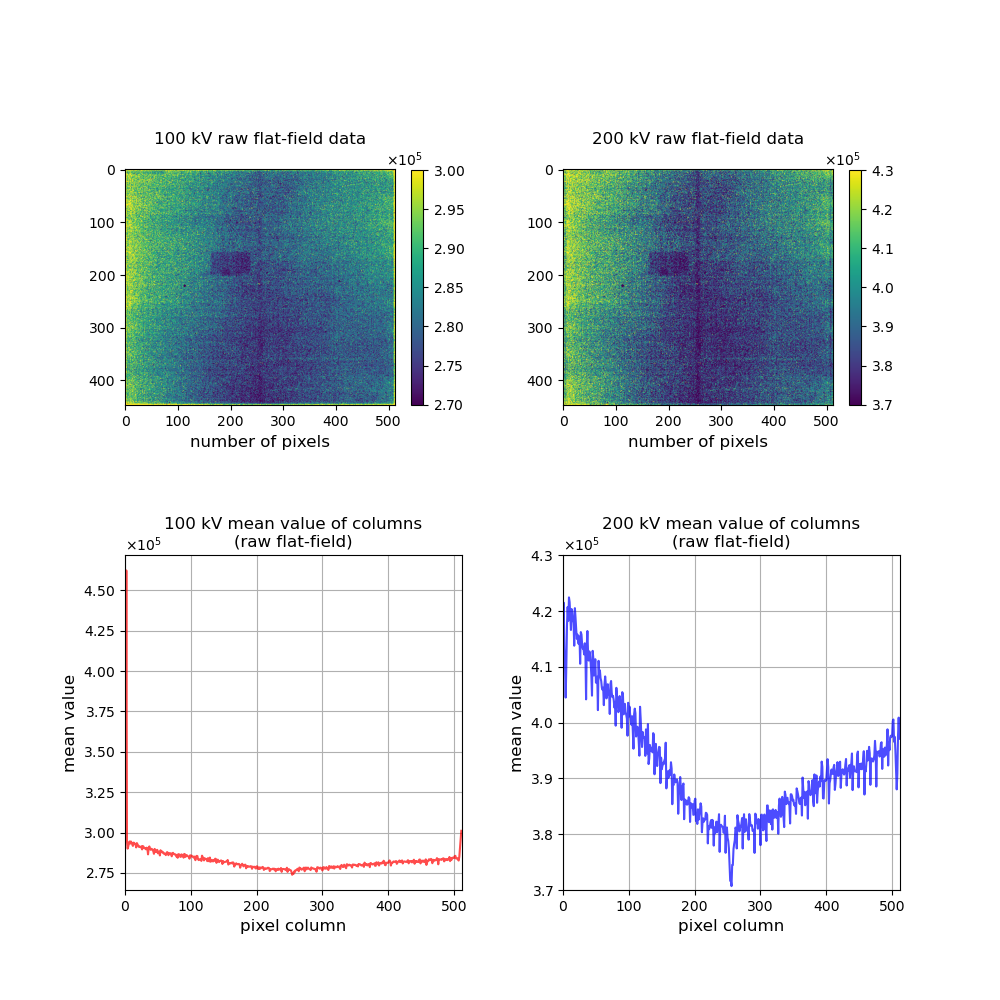}
	\caption{Raw flat-field data at 100~kV and 200~kV with the mean value of the pixel columns.}
	\label{figS:raw_flat_field}
\end{figure}

The original flat-field data (covering the entire detector, 448×512, including the events that were discarded in Section 2.2 because they exceeded $N_{tar}$) show a lower number of events in the central region, as illustrated in Fig~\ref{figS:raw_flat_field}. Across the whole detector, the total number of events at the left boundary is approximately 7\% (100 kV) or 12\% (200 kV) higher than in the central region, while the total number of events at the right boundary is about 3\% (100 kV) or 8\% (200 kV) higher than in the central region, as shown in Fig~\ref{figS:raw_flat_field}. At 100 kV, the 2 edge columns shows abnormal high counts, almost twice the average counts at the left edge column. Therefore, the near edge area is not considered in the NPS calculation.

\begin{figure}[htbp]
	\centering
	\includegraphics[width=1.0\textwidth]{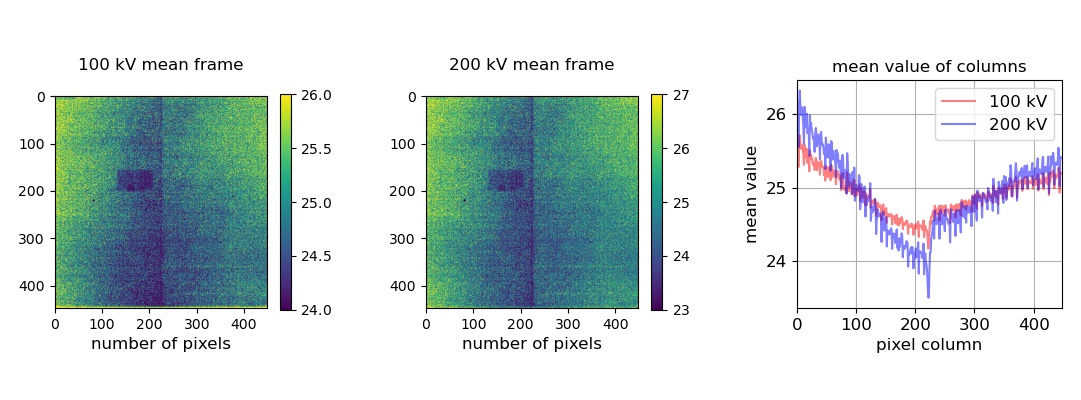}
	\caption{Mean frames recorded at 100~kV and 200~kV, with the mean value of the pixel columns.}
	\label{figS:mean_frame}
\end{figure}

For NPS calulation, although a mean frame (448x448), as shown in Fig~\ref{figS:mean_frame}, is subtracted when calculating the NPS (see section NPS calculation). This low-frequency unevenness (as shown in Fig~\ref{figS:mean_frame}, plot of mean value of columns) may result in an overestimation of the low-frequency component of the NPS. However, in actual imaging experiments, this unevenness can be compensated by flat-field corrections.

\section{Cluster size}
\label{secS:cluster_size}

\begin{figure}[htbp]
	\centering
	\includegraphics[width=1.0\textwidth]{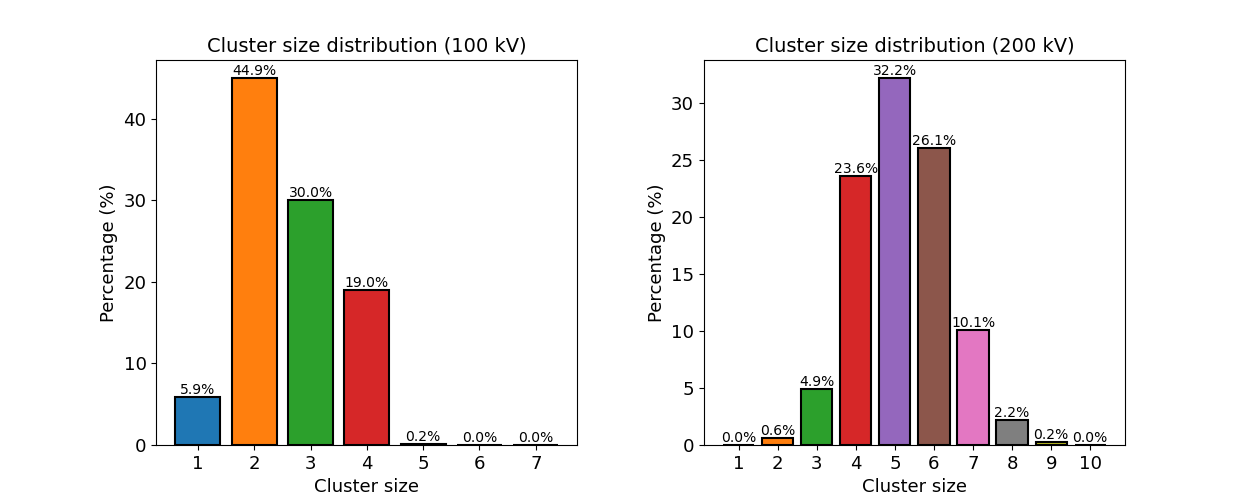}
	\caption{Cluster size distribution at 100 and 200 kV.}
	\label{figS:cluster_size_100kV_200kV}
\end{figure}

An analysis of cluster size distribution is shown in Fig~\ref{figS:cluster_size_100kV_200kV}. The cluster size distributions at 100 kV and 200 kV processed from MTF datasets described in Section 2.2.1, with a per-pixel threshold of 1000 electron-hole pairs (3.6~keV). After clustering~\cite{dimova2025measurement}, only well-reconstructed single-electron responses, excluding partially reconstructed and merged events are counted in the Fig~\ref{figS:cluster_size_100kV_200kV}. The variance of cluster size is calculated from the cluster distribution as 0.746 at 100 kV and 1.361 at 200 kV.

%\bibliographystyle{unsrt}
%\bibliography{bibliography/ref} 

\renewbibmacro{in:}{}
\renewcommand*{\newunitpunct}{\addcomma\space}

\DeclareFieldFormat[article]{volume}{\textbf{#1}}
\DeclareFieldFormat[article]{number}{(#1)}
\DeclareFieldFormat[article]{pages}{#1}
\printbibliography

\end{document}